\documentstyle[prd,aps,epsfig,amssymb]{revtex}
\let\jnfont=\rm
\def\NPB#1,{{\jnfont Nucl.\ Phys.\ B }{\bf #1},}
\def\PLB#1,{{\jnfont Phys.\ Lett.\ B }{\bf #1},}
\def\EPJC#1,{{\jnfont Euro.\ Phys.\ J.\ C }{\bf #1},}
\def\PRD#1,{{\jnfont Phys.\ Rev.\ D }{\bf #1},}
\def\PRL#1,{{\jnfont Phys.\ Rev.\ Lett.\ }{\bf #1},}
\def\MPLA#1,{{\jnfont Mod.\ Phys.\ Lett.\ A }{\bf #1},}
\def\JPG#1,{{\jnfont J.\ Phys.\ G}{\bf #1},}
\def\CTP#1,{{\jnfont Commun.\ Theor.\ Phys.\ }{\bf #1},}

\def\p_slash{\not{\hbox{\kern-2.1pt $p$}}}
\def\k_slash{\not{\hbox{\kern-2.1pt $k$}}}
\begin{document}
\draft
\preprint{}

\title{ Probing New Physics from Top-charm Associated Productions at Linear Colliders}

\author{Junjie Cao $^{1,2}$, Guoli Liu $^2$,  Jin Min Yang $^{3,2}$ \\\ }

\address{$^1$ Department of Physics, Henan Normal University, Henan 453002, China}
\address{$^2$ Institute of Theoretical Physics, Academia Sinica, Beijing 100080, China}
\address{$^3$ CCAST (World Laboratory), P. O. Box 8730, Beijing 100080, China}
\date{\today}

\maketitle

\begin{abstract}
The top-charm associated productions via $e^+ e^-$,  $e^- \gamma$
and $\gamma \gamma$ collisions at linear colliders, which are
extremely suppressed in the Standard Model (SM), could be
significantly enhanced in some extensions of the SM.  In this
article we calculate the full contribution of the
topcolor-assisted technicolor (TC2) to these productions  and then
compare the results with the existing predictions of the SM, the
general two-Higgs-doublet model and the Minimal Supersymmetric
Model. We find that the TC2 model predicts much larger production
rates than other models and the largest-rate channel is $\gamma
\gamma \to t \bar{c}$, which exceeds 10 fb for a large part of the
parameter space. From the analysis of the observability of such
productions at the future linear colliders, we find that
the predictions of the TC2 model can reach the observable level
for a large part of the parameter space while the predictions of other
models are hardly accessible.

\end{abstract}

\section{Introduction}
Since the measurements of top quark properties in Run~I at the
Fermilab Tevatron have small statistics, there remains plenty of
room for new physics in the top quark sector. This stimulates a
lot of efforts in the study of the top quark as a probe of new
physics. Theoretical studies show that the top quark processes are
sensitive to new physics \cite{sensitive}. In some new physics
models like the popular Minimal Supersymmetric Model (MSSM) and
the topcolor-assisted technicolor (TC2) model
\cite{tc2-Hill,tc2-Lane,tc2-Cvetic}, the top quark may have some
exotic production and decay channels
\cite{new-production,new-decay,hjhe,top-Higgs,pp2tc,tc-production,tctcv,tcv,eetc-mssm,ma1,ma2,eetc-tc2,ee-ht,rr-tc}.
Among these exotic processes, one kind is induced by the
flavor-changing neutral-current (FCNC) interactions, which are
extremely suppressed in the SM but could be significantly enhanced
in some extensions
\cite{hjhe,top-Higgs,pp2tc,tc-production,tctcv,tcv,eetc-mssm,ma1,ma2,eetc-tc2,ee-ht,rr-tc}.
Searching for these exotic processes will serve as a good probe
for new physics. Now such possibilities exist on our horizon: the
ongoing Fermilab Tevatron collider, the upcoming CERN Large Hadron
Collider (LHC) and the planned International Linear Collider
(ILC) will allow to scrutinize the top quark nature \cite{review}.

Due to its rather clean environment, the ILC will be an ideal
machine to probe new physics. In such a collider, in addition to
$e^+ e^-$ collision, we can also realize $\gamma \gamma$ collision
and $e^- \gamma$ collision with the photon beams generated by the
backward Compton scattering of incident electron and laser beams
\cite{JLC}. The FCNC top-charm associated productions via $e^+
e^-$,  $e^-\gamma$ and $\gamma \gamma$ collisions will be a
sensitive probe for different new physics models and should be
seriously examined. While these processes have been studied
thoroughly in the MSSM \cite{eetc-mssm}, the corresponding studies
in the TC2 model are not complete: so far only the production at
$e^+e^-$ collision has been studied \cite{eetc-tc2,ee-ht}. From
the studies of these processes in the MSSM~\cite{eetc-mssm} we
know that the $e^+ e^-$ collision channel has a much smaller rate
than $\gamma \gamma$ collision or $e^- \gamma$ collision. So it is
necessary to consider all the production channels to complete the
calculations in TC2 models. This is one aim of this article.

The other aim of this article is to compare the $t\bar c$
production rates predicted by different new physics models.  Such
analysis will help to distinguish different models once the
production rate is measured at the ILC.

The reason for examining the TC2 effects in such top quark
processes is two-fold. One is that the TC2 model is a popular
realization of the fancy idea of technicolor and remains a typical
candidate for new physics in the direction of dynamical symmetry
breaking. This model has not been excluded by experiments so far
and will face the test at future collider experiments.  The other
reason is that the TC2 model may have richer top-quark
phenomenology than other new physics models since it treats the
top quark differently from other fermions. In fact,  the TC2 model
predicts some anomalous couplings for the top quark, among which
the most notable ones occur in the flavor-changing
sector\cite{hjhe}. So the TC2 model may predict a large top-charm
associated production rate and hence single it out from other new
physics models.

This paper is organized as follows. Sec. II presents the
calculations of the top-charm associated productions in TC2 model
at the ILC. Sec. III compares the TC2 results with the
predictions of the two-Higgs-doublet model and the MSSM.
A discussion on
observability at the ILC is given in Sec. IV and the conclusion is
given in Sec. V.

\section{Top-charm associated productions in TC2 model}
\subsection{The relevant Lagrangian}

Among various kinds of dynamical electroweak symmetry breaking
models,  the TC2 model \cite{tc2-Hill,tc2-Lane,tc2-Cvetic} is
especially attractive since it connects the top quark with the
electroweak symmetry breaking (EWSB). In this model, the topcolor
interactions make small contributions to the EWSB, but give rise
to the main part of the top quark mass $(1-\epsilon)m_{t}$ with a
model dependent parameter $\epsilon$. The technicolor interactions
play a main role in the EWSB, and the extend technicolor (ETC)
interactions generate masses of lighter fermions and give
contribution $\epsilon m_{t}$ to the full $m_t$. This model
predicts some scalars such as top-pions ($\pi_{t}^{0},
\pi_{t}^{\pm}$), which are condensates of the third generation
quarks and have strong couplings with the third generation quarks.
The existence of these new particles can be regarded as a typical
feature of the TC2 model. Another feature of the TC2 model is the
existence of large flavor-changing couplings\cite{hjhe,top-Higgs}.
In TC2 models the topcolor interactions are non-universal and
therefore do not posses a Glashow-Illiopoulos-Maiani (GIM)
mechanism. This non-universal gauge interactions result in some
FCNC vertices when one writes the interactions in the quark mass
eigenbasis. Furthermore, the neutral scalars of TC2 model as $t
\bar{t} $ condensates also exhibits the same FCNC vertices.  For
instance, the interactions of top-pions take the
form\cite{tc2-Hill,hjhe}
\begin{eqnarray}
\frac{m_{t}}{\sqrt{2}F_{t}}\frac{\sqrt{v_{w}^{2}-F_{t}^{2}}}
{v_{w}}&[&iK_{UR}^{tt}K_{UL}^{tt*}\bar{t}_L t_{R}\pi_{t}^{0}+
\sqrt{2}K_{UR}^{tt*}K_{DL}^{bb}\bar{t}_R b_{L}\pi_{t}^{+}+
iK_{UR}^{tc}K_{UL}^{tt*}\bar{t}_L c_{R}\pi_{t}^{0}+
\sqrt{2}K_{UR}^{tc*}K_{DL}^{bb}\bar{c}_R b_{L}\pi_{t}^{+}+h.c.],
\end{eqnarray}
where $F_{t}$ is the top-pion decay constant, $v_{w}\equiv
v/\sqrt{2} \simeq 174$ GeV, and $K_{UL} $, $ K_{DL} $ and $ K_{UR}
$ are the rotation matrices that transform the weak eigenstates of
left-handed up-type, down-type and right-handed up-type quarks to
the mass eigenstates, respectively.

In TC2 model, both the up-type and down-type quark mass matrices
($M_U$ and $ M_D $) exhibit an approximate triangle texture at
EWSB scale due to the generic topcolor breaking
pattern\cite{tc2-Hill,Buchalla}, which can severely restrict the
forms of the matrices $K$s in Eq.(1) and in turn, may lead to
contradiction with low energy data. Such a problem was addressed
in \cite{hjhe} and it was observed that, given the textures of the
mass matrices, it is possible to find a natural solution of the
$K$s to evade all the low energy constraints. In this solution, a
realistic but simple pattern of $K_{UL} $ and $ K_{DL} $ is
constructed so that the measured CKM elements are reproduced, and
further, the form of $K_{UR} $, the information of which is hidden
in the SM, can be obtained using the expressions of $M_U $ and
$K_{UL} $. One distinctive character of $K_{UR} $ is that the
mixing between $ t_R $ and $ c_R $ can be naturally large,
reaching $30\% $. This is what we are interested in. Values of the
elements of the $K$s relevant to our discussion are\cite{hjhe}
\begin{equation}
K_{UL}^{tt} \simeq K_{DL}^{bb} \simeq 1, \hspace{5mm}
K_{UR}^{tt}\simeq \frac{m_t^\prime}{m_t} = 1-\epsilon,
\hspace{5mm} K_{UR}^{tc}\leq \sqrt{1-K_{UR}^{tt\ 2}}
=\sqrt{2\epsilon-\epsilon^{2}}, \label{FCSI}
\end{equation}
with $ m_t^\prime $ denoting the topcolor contribution to the top
quark mass. The TC2 model also predicts a CP-even scalar $h_{t}$,
called top-Higgs \cite{top-Higgs}, which is a $\bar{t} t$ bound
state and analogous to the $\sigma$ particle in low energy QCD.
Its couplings to quarks are similar to that of the neutral
top-pion except that the top-Higgs is CP-even while the neutral
top-pion is CP-odd.

\subsection{Analytical calculations}

Since the TC2 contribution to the process $e^+e^-\to  \gamma^\ast,
Z^\ast \to t \bar c$ in $e^+e^-$ collision has already been
calculated in the literature \cite{eetc-tc2,ee-ht},
we focus on the processes $\gamma\gamma \to t\bar c$ in
$\gamma\gamma$ collision and $e^- \gamma \to e^- t\bar c$ in $e \gamma$
collision. For completeness we will also take into account the
process $e^+ e^- \to e^+ e^- \gamma^\ast \gamma^\ast \to e^+ e^-
t\bar{c}$ for $e^+e^-$ collision,  where the $\gamma^{\ast}$
particles are radiated out from the $e^-$ and $e^+$ beams.
Although it is a high-order process compared with $e^+e^-\to
\gamma^\ast, Z^\ast \to t \bar c$, its contributions may not be
underestimated since there is no $s$-channel suppression.

\begin{figure}[]
\vspace*{-1cm}
\hspace*{1cm}
\epsfig{file=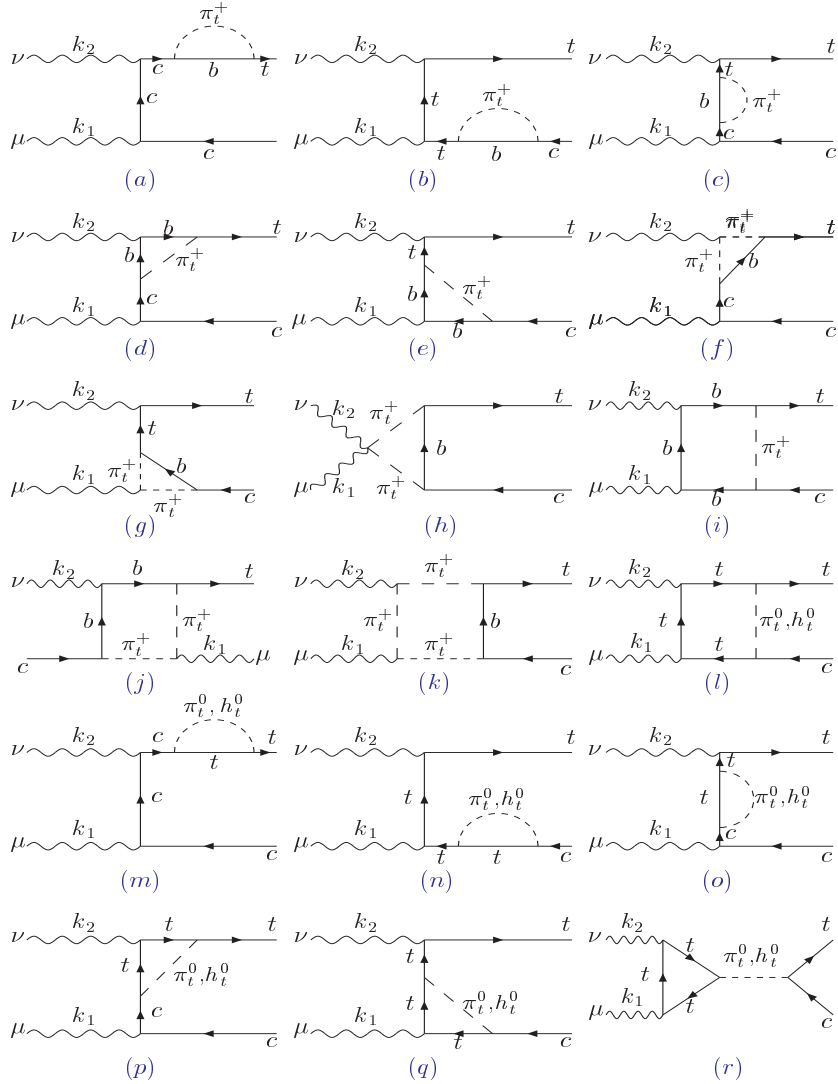,width=18cm,height=21cm}
\vspace*{-5.5cm}
\caption{ Feynman diagrams contributing to the process
$\gamma\gamma \to t\bar c$ in the TC2 model. Those obtained by
exchanging the two external photon lines are not displayed here.}
\label{fig:fig1}
\end{figure}

In TC2 model, the process $\gamma\gamma \to t\bar c$ proceeds at
loop-level by exchanging the top-pions or top-Higgs. The
corresponding Feynman diagrams are shown in Fig.~\ref{fig:fig1}.
Compared with the corresponding diagrams in the
MSSM\cite{eetc-mssm}, the contributions of TC2 model involve
additional $s$-channel contributions, as shown in
Fig.~\ref{fig:fig1}(r). The effective coupling between top-pion
(top-Higgs) and photons can be written as:
\begin{eqnarray}
&& \gamma \gamma \pi_t^0:~~  \frac{2 \alpha_e Q_t^2}{\pi}
\frac{m_t}{\sqrt{2} F_t} \frac{\sqrt{v_w^2-F_t^2}}{v_w}
\frac{m_t}{s} K_{UR}^{tt} K_{UL}^{tt \ast} c_1(\frac{s}{m_t^2}) (
i \epsilon_{\mu \nu \rho
\lambda} k_1^\rho k_2^\lambda )~ \varepsilon_{\mu} (k_1, \lambda_1) \varepsilon_{\nu} (k_2, \lambda_2),   \\
&& \gamma \gamma h_t:~~ \frac{2 \alpha_e Q_t^2}{\pi}
\frac{m_t}{\sqrt{2} F_t} \frac{\sqrt{v_w^2-F_t^2}}{v_w}
\frac{m_t}{s} K_{UR}^{tt}
 K_{UL}^{tt \ast} c_2(\frac{s}{m_t^2}) (\frac{s}{2} g_{\mu \nu} - k_{1 \nu}
 k_{2 \mu})~ \varepsilon_{\mu} (k_1, \lambda_1) \varepsilon_{\nu} (k_2, \lambda_2),
\end{eqnarray}
where  $s=2 k_1\cdot k_2$, $c_1 (R)= \int_0^1 {\it d} x \frac{\ln [1-R x (1-x)]}{x}$,
$ c_2 (R)= -2 +\left( 1 -\frac{4}{R} \right ) c_1 (R)$,
$Q_t $ is the electric charge of the top quark,
$k_{1,2}$ denote the momentum of the photons,
and $\varepsilon_{\nu}(k_{1,2}, \lambda_{1,2})$ are the polarization
vectors of the photons.
 For the top-pion
(top-Higgs) within the range between $ m_t $ and $ 2 m_t$, the
top-pion (top-Higgs) decays dominantly into $t \bar{c}
$\cite{hjhe,top-Higgs} \footnote{Depending on their masses, the
decay products of neutral top-pion and top-Higgs may be $t \bar{t}
$, $t \bar{c} $, $b \bar{b} $, $g g $,$ W^+ W^- $, $ Z^0 Z^0 $,$
\gamma \gamma $ and $ Z^0 \gamma $. Generally speaking, the
dominant decay mode is $t \bar{t} $ for $m_{\pi_t, h_t} > 2 m_t $,
$t \bar{c} $ for $ m_t < m_{\pi_t, h_t} < 2 m_t $ and  $ b \bar{b}
$ for $ m_{\pi_t, h_t} < m_t $.} and thus such
$s$-channel contributions become dominant if the c.m. energy of
$\gamma \gamma $ collision is high enough to produce a real
top-pion (top-Higgs). In this case the cross section of top-charm
associated production in the TC2 model may be quite large. In our
calculation, we have considered all the decay channels of the
top-pion and top-Higgs and taken into account the width effects in
the $s$-channel propagators.

Note that the reliability for doing such perturbative calculations
should be carefully checked since the Yukawa couplings between the
top-pion and the top quark  might be large in TC2 model. For
example, from Eq.(1) we see that such coupling strength is
$Y=\frac{m_{t}}{\sqrt{2}F_{t}}\frac{\sqrt{v_{w}^{2}-F_{t}^{2}}}
{v_{w}}$, which yields $Y^2/4 \pi \simeq 0.44$ for a typical value
of $F_t $ ( $F_t=50 $ GeV ), a large value but still making the
perturbative expansion valid. One may then wonder whether the
perturbative expansion involving such large Yukawa couplings
converge well, or, in other words, whether the TC2 Yukawa
corrections to top quark processes can drastically change the
leading order predictions. Actually, such large Yukawa couplings
are often present in new physics models such as the two-Higgs
doublet model \cite{2hdm} and the supersymmetric model \cite{MSSM}
with small or large $\tan \beta $. One-loop Yukawa corrections to
top production processes at hadron and linear colliders have been
calculated \cite{yukawa-correction}, and it was observed that the
corrections can maximally reach $30 \%$ in amplitude and drop
rapidly as the scalars becoming heavy.  So we believe that the
perturbative expansion involving the TC2 Yukawa couplings is still
a good expansion.

Another point that should be addressed is whether the top-pions
can be regarded as point-like particles at the ILC energy. In the
TC2 model the strong top-color gauge interaction causes the
top-quark to condensate, which leads to the presence of top-pions.
The compositeness scale of top-pions is about the mass scale of
top-color gauge boson, which is usually assumed to be a few
TeV\cite{tc2-Hill,tc2-Cvetic}. So at the ILC energy (a few hundred
GeV to 1 TeV), the top-pions can be regarded as point-like
particles. Theoretically, the masses of top-pions can be quite
light, well below the top-color scale, since they are a kind of
pseudo Goldstone bosons (a detailed discussion on their masses are
given in the proceeding section)\cite{tc2-Hill}. So the top-pions
may be accessible at the ILC although the top-color scale is well
above the ILC energy.

The magnitude of $\gamma\gamma \to t\bar c$ can be written as
\begin{eqnarray}
{\cal M} &=&\frac{\alpha_e m_t^2}{4 \pi
F_t^2}\frac{v_w^2-F_t^2}{v_w^2} K_{UR}^{tt*} K_{UR}^{tc}
 \sum_i \bar{u}_t ~\Gamma_{i}^{\mu\nu } \frac{1+\gamma_5}{2} ~v_c
 \varepsilon_{\mu} (k_1, \lambda_1) \varepsilon_{\nu} (k_2,
 \lambda_2) , \label{expression}
\end{eqnarray}
where the sum is over all Feynman diagrams and for each diagram
$F_i^{\mu \nu}$ takes the form
\begin{eqnarray}
\Gamma^{\mu\nu}_i&=&
    c_{i,1}p_t^{\mu}p_t^{\nu}+c_{i,2}p_c^{\mu}p_c^{\nu}+c_{i,3}p_t^{\mu}p_c^{\nu}
   +c_{i,4}p_t^{\nu}p_c^{\mu} +c_{i,5}p_t^{\mu}\gamma^\nu+c_{i,6}p_c^{\mu}\gamma^\nu
   +c_{i,7}p_c^{\nu}\gamma^\mu+c_{i,8}p_t^{\nu}\gamma^\mu
+c_{i,9}g^{\mu\nu} +c_{i,10}\gamma^{\nu}\gamma^{\mu} \nonumber \\
&&+c_{i,11}p_t^{\mu}p_t^{\nu} \k_slash_2
   +c_{i,12}p_c^{\mu}p_c^{\nu}\k_slash_2 +c_{i,13}p_t^{\mu}p_c^{\nu} \k_slash_2
   +c_{i,14}p_t^{\nu}p_c^{\mu} \k_slash_2
  +c_{i,15}p_t^{\mu}\gamma^\nu \k_slash_2 +c_{i,16}p_c^{\mu}\gamma^\nu
  \k_slash_2 \nonumber \\ &&
+c_{i,17}p_c^{\nu}\gamma^\mu \k_slash_2
   +c_{i,18}p_t^{\nu}\gamma^\mu \k_slash_2  +c_{i,19}g^{\mu\nu} \k_slash_2
   +c_{i,20}i\varepsilon^{\mu\nu\alpha\beta}\gamma_\alpha k_{2\beta} +
   c_{i,21} k_{1 \mu} p_{t \nu}+ c_{i,22} k_{1 \mu} p_{c \nu } \nonumber \\
   && +c_{i,23} k_{2 \nu} p_{t \mu} + c_{i,24} k_{2 \nu} p_{c \mu }+
   c_{i,25} k_{1 \mu} \gamma_{\nu}+ c_{i,26} k_{2 \nu }
   \gamma_{\mu} +c_{i,27} k_{1 \mu} p_{t \nu}\k_slash_2 + c_{i,28} k_{1 \mu} p_{c \nu }\k_slash_2 \nonumber \\&&
   +c_{i,29} k_{2 \nu} p_{t \mu}\k_slash_2 + c_{i,30} k_{2 \nu} p_{c
\mu }\k_slash_2+ c_{i,31} k_{1 \mu} \gamma_{\nu}\k_slash_2+
c_{i,32} k_{2 \nu }
   \gamma_{\mu}\k_slash_2. \label{m-e}
\end{eqnarray}
Here, $p_{t,c}$ are the momentum of outgoing top and charm quarks. The
coefficients $c_{i,j}$ ($ j=1 $, $ \cdots 32 $)  can be obtained by a
straightforward calculation of each Feynman diagram. For the sake
of conciseness, we do not present their lengthy expressions here.

We checked that all ultraviolet divergences cancel out in our
results, which is essentially guaranteed by the renormalizability.
We also checked that our results satisfy the Ward identity,
$k_1^\mu\Gamma_{\mu\nu}=0 $ and $ k_2^\nu \Gamma_{\mu\nu}=0$ with
$ \Gamma_{\mu \nu} $ being the sum of  $\Gamma_{i \mu \nu} $. In
fact, the $s$-channel contribution mediated by either top-pion or
top-Higgs (shown in Fig.\ref{fig:fig1} (r)) satisfies the Ward
identity separately. This fact enables one to consider the
$s$-channel contribution separately, as in Ref\cite{rr-tc}. For
the convenience of our discussion, we will divide TC2
contributions into the $s$-channel contribution and the non
$s$-channel contribution, the latter of which includes the
contributions from the $t$-channel, $u$-channel, quadric as well
as box diagrams.

The  $t \bar c $ production in  $e^- \gamma$ collision proceeds
through the process $e^- \gamma \to e^- \gamma^{\ast} \gamma  \to
e^- t \bar c $, where the $\gamma$-beam is generated by the
backward Compton scattering of incident electron- and laser-beam
and the $\gamma^{\ast}$ is radiated from $e^-$ beam. The
subprocess $\gamma^{\ast} \gamma  \to t \bar c $ has the same
Feynman diagrams as those shown in  Fig.~\ref{fig:fig1}. In our
calculation we use the Weizs$\ddot{a}$cker-Williams approximation
\cite{willam} which treats $\gamma^{\ast}$ from $e^-$ beam as a
real photon. Thus the cross section is given by
\begin{eqnarray}
\hat \sigma_{e^- \gamma \to  e^- t \bar c}(s_{e \gamma})&=&
\int_{(m_t+m_c)^2/s_{e\gamma}}^1 {\rm d} x~P_{\gamma/e}(x,E_e)
~\hat{\sigma}_{\gamma \gamma \to  t \bar c}(s_{\gamma \gamma}=x
s_{e \gamma}) , \label{e-gamma}
\end{eqnarray}
where $P_{\gamma/e} (x, E_e)$ is the probability of finding a
photon with a fraction $x$ of energy $E_e$ in an
ultra-relativistic electron and is given by \cite{willam}
\begin{eqnarray}
P_{\gamma/e}(x, E_e)=\frac{\alpha}{\pi} \left (
\frac{1+(1-x)^2}{x}
 \left ( \ln{\frac{E_e}{m_e}}-\frac{1}{2} \right )+
\frac{x}{2} \left ( \ln{(\frac{2}{x}-2)}+1 \right ) +
\frac{(2-x)^2}{2 x} \ln{\left ( \frac{2-2 x}{2-x} \right )}
\right) .
\end{eqnarray}
Note that the incoming electron
may also radiate $Z$-boson to contribute to the process $e^- \gamma
\to e^- t \bar c$. However, such contribution is suppressed by the
probability function of finding a $Z$-boson in an
ultra-relativistic electron \cite{effield} and can be safely
neglected.

As for $e^+ e^- \to e^+ e^- \gamma^\ast \gamma^\ast \to e^+ e^- t
\bar{c}$ in  $e^+ e^-$ collision, its cross section can be
obtained by folding $ P_{\gamma/e} $ with $ \hat{\sigma}_{\gamma
\gamma \to t \bar{c}} $, as done in Eq.(\ref{e-gamma}),
\begin{eqnarray}
\sigma_{e^+ e^- \stackrel{\gamma^\ast \gamma^\ast}{\to}  e^+ e^- t
\bar c}(s_{e^+ e^-})&=& \int_a^1 {\rm d} x \int_{a/x}^1 {\rm d} y
P_{\gamma/e}(x,E_{e^+}) P_{\gamma/e} (y, E_{e^-})
~\hat{\sigma}_{\gamma \gamma \to t \bar
c}(s_{\gamma \gamma}=x  y s_{e^+e^-}) \nonumber \\
&=& \int_{\sqrt{a}}^1 2 z {\rm d} z ~\hat{\sigma}_{\gamma \gamma
\to t \bar c}
 (s_{\gamma \gamma}= z^2 s_{e^+e^-})
\int_{z^2}^{1} \frac{{\rm d} x}{x}
 ~P_{\gamma/e}(x, E_{e^+}) P_{\gamma/e}(\frac{z^2}{x}, E_{e^-}).
 \label{cross2}
\end{eqnarray}
where we define $a=(m_t+m_c)^2/s_{e^+e^-}$. In TC2 model, it is argued
that the top-Higgs may couple directly with $ZZ$ or
$W^+W^-$\cite{hjhe}. If so, the processes $e^+ e^- \to e^+ e^- ZZ
\to e^+ e^- t \bar{c} $ and $e^+ e^- \to \nu \bar{\nu} W^+ W^- \to
\nu \bar{\nu} t \bar{c}$ may also be important for top-charm
associated production at $e^+ e^-$ collision \cite{ee-ht}. We will
take the results of Ref.\cite{ee-ht} for comparison in our
discussions.

For both $\gamma \gamma$ collision and $e \gamma$ collision, the
photon beams are generated by the backward Compton scattering of
incident electron- and laser-beams just before the interaction
point. The events number is obtained by convoluting the cross
section with the photon beam luminosity distribution. For $\gamma
\gamma$ collider the events number is obtained by
\begin{eqnarray}
N_{\gamma \gamma \to  t \bar c}&=&\int  {\rm d} \sqrt{s_{\gamma
\gamma}} \frac{{\rm d}{\cal L}_{\gamma \gamma}} {{\rm d}
\sqrt{s_{\gamma \gamma}}} \hat{\sigma}_{\gamma \gamma \to  t \bar
c} (s_{\gamma \gamma})\equiv {\cal L}_{e^+e^-} ~\sigma_{\gamma \gamma
\to  t \bar c}(s_{e^+e^-}), \label{definition}
\end{eqnarray}
where ${\rm d}{\cal L}_{\gamma \gamma}/{\rm d} \sqrt{s_{\gamma
\gamma}}$ is the photon beam luminosity distribution and
$\sigma_{\gamma \gamma \to  t \bar c}(s_{e^+e^-}) $, with $s_{e^+e^-}$
being the energy-square of $e^+e^-$ collision, is defined as the
effective cross section of $ \gamma \gamma \to  t \bar c$.  In
optimum case, it can be written as \cite{rrcl}
\begin{eqnarray}
\sigma_{\gamma \gamma \to  t \bar
c}(s_{e^+e^-})&=&\int_{\sqrt{a}}^{x_{max}} 2 z{\rm d} z
 ~\hat{\sigma}_{\gamma \gamma \to  t \bar c} (s_{\gamma \gamma}=z^2 s_{e^+e^-})
\int_{z^2/x_{max}}^{x_{max}} \frac{{\rm d} x}{x}~F_{\gamma/e}(x)
~F_{\gamma/e}(\frac{z^2}{x}), \label{cross}
\end{eqnarray}
where $F_{\gamma/e}$ denotes the energy spectrum of the
back-scattered photon for unpolarized initial electron and laser
photon beams given by
\begin{eqnarray}
F_{\gamma/e}(x)&=&\frac{1}{D(\xi)} \left (
1-x+\frac{1}{1-x}-\frac{4 x}{\xi (1-x)}+ \frac{4 x^2}{\xi^2
(1-x)^2} \right ) .
\end{eqnarray}
The definitions of parameters $\xi$, $D(\xi)$ and $x_{max}$ can be
found in \cite{rrcl}. In our numerical calculation, we choose
$\xi=4.8$, $D(\xi)=1.83$ and $x_{max}=0.83$.

For the $e^- \gamma $ collider the effective cross section of $e
\gamma \to  e t \bar c$ is defined as
\begin{eqnarray}
\sigma_{e^- \gamma \to  e^- t \bar c}(s_{e^+e^-})&=& \frac{1}{{\cal
L}_{e^+e^-}} \int {\rm d} \sqrt{s_{e\gamma}} \frac{{\rm d}{\cal L}_{e
\gamma}}
{{\rm d} \sqrt{s_{e \gamma}}} ~\hat \sigma_{e^- \gamma \to  e^- t \bar c} (s_{e \gamma}) \nonumber \\
&=& \int_{\sqrt{a}}^{\sqrt{x_{max}}} 2 z{\rm d} z
~\hat{\sigma}_{\gamma \gamma \to  t \bar c}
 (s_{\gamma \gamma}=z^2 s_{e^+e^-})
\int_{z^2/x_{max}}^{1} \frac{{\rm d} x}{x}
 ~P_{\gamma/e}(x, E_e) F_{\gamma/e}(\frac{z^2}{x}).
\label{cross1}
\end{eqnarray}

From above analysis, especially from
Eqs.(\ref{cross2},\ref{cross},\ref{cross1}), we know the cross
sections for $e^+ e^- \stackrel{\gamma^\ast \gamma^\ast}{\to} e^+
e^- t \bar{c} $, $ e^- \gamma \to e^- t \bar{c} $ and $ \gamma
\gamma \to t \bar{c} $ are connected by convoluting the same
$\hat{\sigma}_{\gamma \gamma \to t \bar{c}} $ with different
photon distribution functions. Noting the fact $ F_{\gamma/e} >
P_{\gamma/e} $ for the integrated range of x, one can infer that
the cross section at $\gamma \gamma $ collider is the largest,
which will be shown in our results. In the following the major
part of discussions will be focused on the production in  $\gamma
\gamma$ collision.

\subsection{Numerical results}

In our numerical study, the bottom and charm quark masses will
be neglected, and  the charge conjugate $\bar t c$ production
channel is also included.
The cross sections of top-charm associated productions in TC2
model depend on $\epsilon$, $ K_{UR}^{tc}$, the top-pion decay constant
$F_{t}$ and the masses of the top-pions and top-Higgs. Before
starting numerical calculations, we recapitulate the theoretical and
experimental constraints on these parameters.
\begin{itemize}
\item[{\rm (1)}] About the $\epsilon$ parameter. In TC2 model, $\epsilon $
parameterizes the portion of ETC contribution to the top quark
mass. The bare value of $\epsilon $ is generated at the ETC scale,
and subject to very large radiative enhancement from Topcolor and
$U(1)_{Y_1} $ by a factor of order $10$ when evolving  down to the
weak scale\cite{tc2-Hill}. This $\epsilon $ can induce a nonzero
top-pion mass (proportional to $\sqrt{\epsilon} $ )\cite{Hill} and
thus ameliorate the problem of having dangerously light scalars.
Numerical analysis shows that, with reasonable choice of other
input parameters, $\epsilon $ of order $10^{-2} \sim 10^{-1} $ may
induce top-pions as massive as the top quark\cite{tc2-Hill}.
Indirect phenomenological constraints on $\epsilon $ come from low
energy flavor changing processes such as $ b \to s \gamma $
\cite{b-sgamma}. However, these constraints are very weak. Precise
value of $\epsilon $ may be obtained by elaborately measuring the
coupling strength between top-pion/top-Higgs and tops at the
linear colliders. From the theoretical point of view, $\epsilon $ with
value from $ 0.01 $ to $ 0.1 $ is favored. For the considered
process in our analysis, the non-zero $\epsilon $ contributes a factor
of $(1- \epsilon )^2$ (through $(K_{UR}^{tt})^2$) to the cross section
(see Eq.(\ref{expression})), and thus the results are not sensitive
to $\epsilon$ in the range of  $0.01 \sim 0.1$.
Throughout this paper, we fix conservatively $\epsilon =0.1$.

\item[{\rm (2)}] The parameter $K_{UR}^{tc}$ is upper bounded by
unitary relation $K_{UR}^{tc} \leq \sqrt{1-K_{UR}^{tt\ 2}}
=\sqrt{2\epsilon -\epsilon^2}$. For a $\epsilon $ value smaller
than $0.1 $, this corresponds to $ K_{UR}^{tc} < 0.43$. In our
analysis, we will treat $K_{UR}^{tc}$ as a free parameter.

\item[{\rm (3)}] About the top-pion decay constant $F_t$,  the Pagels-Stokar formula
\cite{Pagels} gives an expression for it in terms of the number of
quark color $N_c$, the top quark mass, and the scale $\Lambda $ at which the
condensation occurs:
\begin{eqnarray}
F_t^2= \frac{N_c}{16 \pi^2} m_t^2 \ln{\frac{\Lambda^2}{m_t^2}}.
\label{ft}
\end{eqnarray}
From this formula, one can infer that, if $t\bar{t} $ condensation
is fully responsible for EWSB, i.e. $F_t \simeq v_w \equiv
v/\sqrt{2} = 174$ GeV, then $\Lambda $ is about $10^{13} \sim
10^{14} $ GeV. Such a large value is less attractive since by the
original idea of Technicolor theory\cite{Farhi}, one expects new
physics scale should not be far higher than the weak scale. On the
other hand, if one believe new physics exists at TeV scale, i.e.
$\Lambda \sim 1 $ TeV, then $F_t \sim 50$ GeV, which means that $t
\bar{t} $ condensation alone cannot be wholly responsible for EWSB
and to break electroweak symmetry needs the joint effort of
Topcolor and other interactions like Technicolor.  From the
experimental point of view, probably the best way to determine
$F_t $ is by precisely measuring the coupling strength of
top-Higgs with vector bosons at future linear collider, which is
proportional to $F_t $ without any theoretical
ambiguity\cite{top-Higgs}. By the way, Eq.(\ref{ft}) should be
understood as only a rough guide, and $F_t $ may in fact be
somewhat lower or higher, say in the range $40 \sim 80$ GeV.
Allowing $F_t $ to vary over this range does not qualitatively
change our conclusion, and, therefore, we use the value $F_t =50$
GeV for illustration in our numerical analysis.

\item[{\rm (4)}] About the mass bounds for top-pions and
top-Higgs. On the theoretical side, some estimates have been done.
The mass splitting between the neutral top-pion and the charged
top-pion should be small since it comes only from the electroweak
interactions\cite{mass-pion}. Ref.\cite{tc2-Hill} has estimated
the mass of top-pions using quark loop approximation and showed
that $m_{\pi_t}$ is allowed to be a few hundred GeV in reasonable
parameter space. Like Eq.(\ref{ft}), such estimations can only be
regarded as a rough guide and the precise values of top-pion
masses can be determined only by future experiments. The mass of
the top-Higgs $h_{t}$ can be estimated in the Nambu-Jona-Lasinio
(NJL) model in the large $N_{c}$ approximation and is found to be
about $2m_{t}$ \cite{top-Higgs}. This estimation is also rather
crude and the mass below the $\overline{t}t$ threshold is quite
possible in a variety of scenarios \cite{y15}. On the experimental
side,  current experiments have restricted the mass of the charged
top-pion. For example, the absence of $t \to \pi_t^+b$ implies
that $m_{\pi_t^+} > 165$ GeV \cite{t-bpion} and $R_b$ analysis
yields $m_{\pi_t^+}> 220$ GeV \cite{burdman,kuang}. For the
neutral top-pion and top-Higgs, the experimental restrictions on
them are rather weak. (Of course, considering theoretically that
the mass splitting between the neutral and charged top-pions is
small, the $R_b$ bound on the charged top-pion mass should be
applicable to the neutral top-pion masses.) The current bound on
techni-pions\cite{datagroup} does not apply here since the
properties of top-pion are quite different from those of
techni-pions. The direct search for the neutral top-pion
(top-Higgs) via $ p p \to t \bar{t} \pi_t^0 (h_t)$ with $\pi_t^0
(h_t) \to b \bar{b} $ was proven to be hopeless at Tevatron for
the top-pion (top-Higgs) heavier than $135 $ GeV \cite{Rainwater}.
The single production of $\pi_t^0 $ ($h_t $ ) at Tevatron with
$\pi_t^0 $ ($h_t $) mainly decaying to $t \bar{c} $ may shed some
light on detecting top-pion (top-Higgs)\cite{top-Higgs}, but the
potential for the detection is limited by the value of
$K_{UR}^{tc}$ and the detailed background analysis is absent now.
Anyhow, these mass bounds will be greatly tightened  at the
upcoming LHC \cite{hjhe,pp2tc,Rainwater}. Combining the above
theoretical and experimental bounds, we will assume
\begin{equation}
m_{\pi_{t}^{0}}=m_{\pi_{t}^+}\equiv m_{\pi_t} > 220 ~{\rm GeV}.
\end{equation}
\end{itemize}

\begin{figure}[]
\begin{center}
\epsfig{file=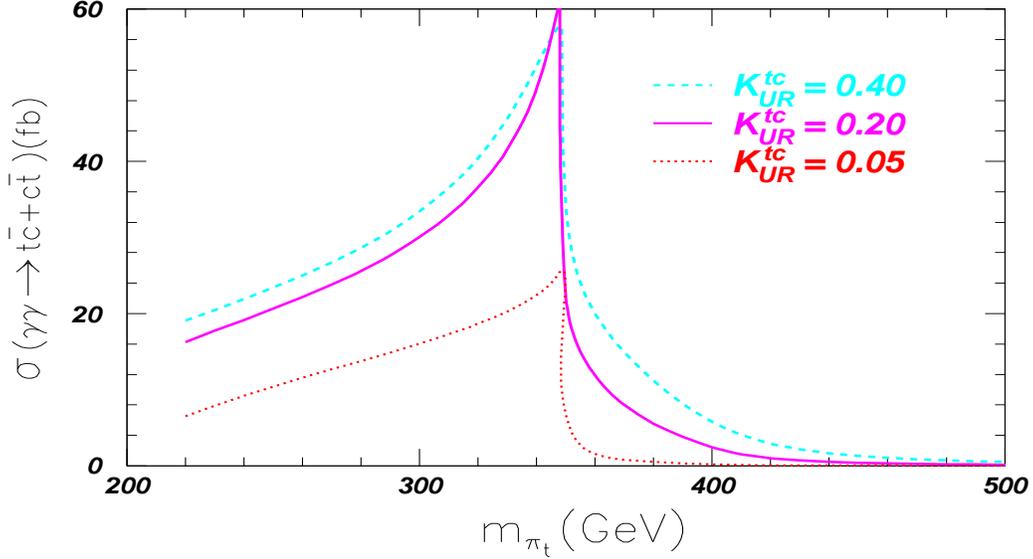,width=14cm,height=8cm} \caption {The $m_{\pi_t}$
dependence of the cross section in $\gamma\gamma$ collision for
$\sqrt{s_{e^+ e^-}} =500$ GeV  under the assumption $m_{\pi_t}=m_{h_t}$.}
\label{fig:fig2}
\end{center}
\end{figure}

Fig.\ref{fig:fig2} shows the dependence of the cross section of
$\gamma \gamma \to t \bar{c} $ on $m_{\pi_t}$ under the assumption
$m_{\pi_t}=m_{h_t}$.
One sees that for $m_{\pi_t} > 220$ GeV, the
cross section first increases monotonously to reach its maximum
value at $m_{\pi_t} =2 m_t$, and then drops rapidly. This behavior
may be explained as follows. For the results in
Fig.\ref{fig:fig2}, the dominant contributions are from the
$s$-channel diagrams (we will comment on the effects of the non
$s$-channel diagrams later), and the cross section then may be
estimated by the narrow width approximation (Note that there is no
interference between the contribution from the top-pion and that
from the top-Higgs due to the different CP property of $\pi_t^0 $
and $ h_t $.):
\begin{eqnarray}
\sigma (\gamma \gamma \to t \bar{c}) =\sigma (\gamma \gamma \to
\pi_t^0 ) Br(\pi_t^0 \to t \bar{c}) + \sigma (\gamma \gamma \to h_t)
Br(h_t \to t \bar{c}) ,
\end{eqnarray}
where $\sigma (\gamma \gamma \to \pi_t^0 (h_t) ) $ is the  rate for
single top-pion (top-Higgs) production, and $Br (\pi_t^0 (h_t) \to t
\bar{c}) $ is the branching fraction for $ \pi_t^0 (h_t) \to t
\bar{c} $. In the range $m_t+m_c < m_{\pi_t} < 2 m_t$, the
cross section increases with $m_{\pi_t}$ since the $t \bar{c}$
decay mode of $\pi^0_t$($h_t$) is getting a larger branching ratio
and becoming dominant as $m_{\pi_t}$ increases.
When $m_{\pi_t}$ passes the threshold of $2 m_t$ and
keeps increasing, the cross section drops quickly since
the $t\bar t$ is becoming the dominant  decay mode
of $\pi^0_t$($h_t$) and $\sigma (\gamma \gamma \to \pi_t^0
(h_t) )$ is getting severely suppressed by the photon luminosity
distribution.

\vspace*{-0.5cm}

Fig.\ref{fig:fig2} shows that the $t\bar c$ production rate in
$\gamma \gamma$ collision can exceed $10$ fb. Comparing with
$e^+ e^- \to t \bar c$ in  $e^+ e^-$ collision, which can only
reach $0.1$ fb  in TC2 model \cite{eetc-tc2} due to the
$s$-channel suppression, one may infer that the $\gamma \gamma$
collision may be much better in probing  TC2 model. We will
elaborate the observability of the top-charm associated production
at the ILC in Sev. IV. Since we only intend to figure out the
typical order of the production rate for such rare processes, we
fix $K_{UR}^{tc}=0.4$ for illustration in our following analysis.

\begin{figure}
\begin{center}
\epsfig{file=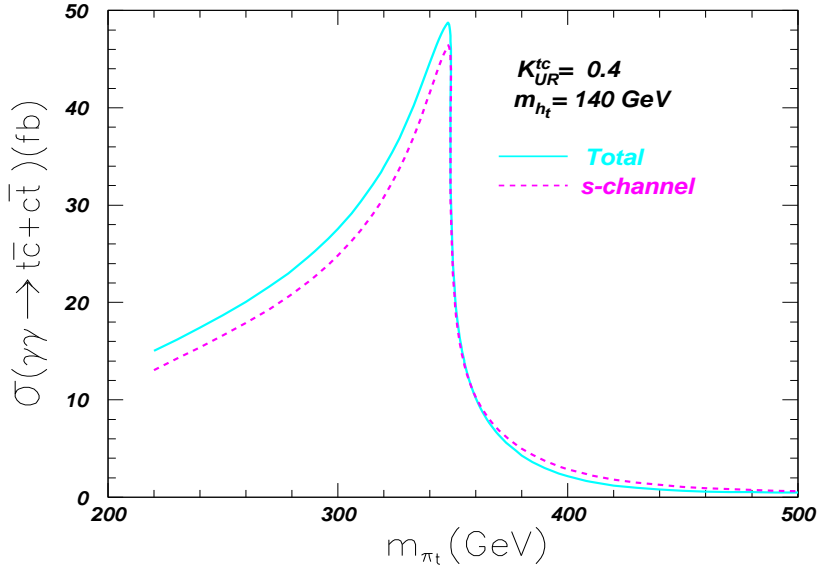, width=11cm,height=8cm}
\caption { The $m_{\pi_t} $ dependence of the cross section of $ \gamma \gamma
\to t \bar{c} $  for $\sqrt{s_{e^+ e^-}}=500$ GeV. The solid curve
is the total contribution from both $s$-channel and non $s$-channel
diagrams, while the dashed is the contribution from only the $s$-channel
diagrams. }
\label{fig:fig3}
\end{center}
\end{figure}

\vspace*{0.2cm}
Now we comment on the effects of the non
$s$-channel diagrams in Fig.\ref{fig:fig1}. There are two
characters for such effects. One is that they are proportional to $K_{UR}^{tc}$,
and, therefore, may be sizable only for large values of $K_{UR}^{tc}$.
The other is that they generally increase with the enhancement of the c.m. energy
of the collider. For $ K_{UR}^{tc} =0.4 $, $\sqrt{s_{e^+e^-}} =500$ GeV and
 $m_{\pi_t}=m_{h_t}$, our findings are:
\begin{itemize}
\item[{\rm (a)}] In the most interesting range $m_t < m_{\pi_t} < 2 m_t$,
  the cross section is dominated by the $s$-channel contributions. The
  non $s$-channel effects, which in this case arise
  mainly from the interference with the $s$-channel diagrams,
  are less than $3\%$.
  The main reason is that, as pointed earlier,
  in this region $t\bar{c}$ is the dominant decay mode of $\pi^0_t$
  and $h_t$ produced in the $s$-channel. Another reason is that under our assumption
  $m_{\pi_t}=m_{h_t}$ there exists a cancellation between the top-pion non $s$-channel
   diagrams and the top-Higgs non $s$-channel diagrams.

\item[{\rm (b)}] In the heavy range $m_{\pi_t} > 2 m_t$, which is less interesting since
  the cross section is too small (as shown in Fig.\ref{fig:fig2}),  the cross section
  is also dominated by  the $s$-channel contributions due to the cancellation of the
  non $s$-channel  top-pion and top-Higgs diagrams mentioned above.

\item[{\rm (c)}] In the light range $m_{\pi_t} < m_t$, which is
disfavored by $R_b$\cite{burdman,kuang} and thus the corresponding
results are not shown here, the non $s$-channel contributions are
dominant. In this region the $s$-channel contributions are
suppressed since the top-pions and the top-Higgs in the $s$-channel
cannot be on-shell. The non $s$-channel contributions are quite large
in this region due to the following two reasons. One is that $u-m_c^2 $
(for $u$-channel) and $t -m_c^2 $ (for $t$-channel) in the
propagators of the charm quark (see Fig.\ref{fig:fig1}
(a,c,d,f,m,p)) can approach zero \footnote{Setting $m_c =0 $ does
not develop poles for the cross section since the poles from the
$u$-channels are canceled by those from the $t$-channels.} and thus
the non $s$-channel contributions can be greatly enhanced. In
fact, this is the advantage of $\gamma \gamma $ collider over $e^+
e^- $ collider in exploring the top quark FCNC processes, as
pointed out in Ref.\cite{effect2}. The other reason is that the
contributions from the box diagrams of Fig.1 (i-k) are also quite
sizable for light top-pions. From our numerical evaluation we
found that in the region $m_{\pi_t} < m_t$, among the non
$s$-channel diagrams, the charged top-pion diagrams give dominant
contributions and the cross section peaks at $ m_{\pi_t} \simeq 140$
GeV with $\sigma_{max}\simeq 25 fb $ .

\item[{\rm (d)}] Note that our above analysis are under the
assumption $m_{h_t}=m_{\pi_t}$.
Although there is a good reason \cite{mass-pion} to expect the
mass degeneracy for neutral and charged top-pions, the top-Higgs
mass $m_{h_t}$ may be quite different from the top-pion mass
$m_{\pi_t}$($=m_{\pi_t^0}=m_{\pi_t^+}$). If we allow a splitting
between $m_{h_t}$ and $m_{\pi_t}$, we found that in the allowed
range of $m_{\pi_t}$ (i.e. $>220$ GeV), the non $s$-channel
contributions can be as large as $16\%$ for a relatively light
top-Higgs, as shown in Fig.\ref{fig:fig3}.
\end{itemize}

So we conclude: (i) Under the assumption $m_{\pi_t}=m_{h_t}$,
the cross section is dominated by the $s$-channel
contributions and the non $s$-channel effects are negligibly small
for $m_{\pi_t} > m_t$. Only in the case of $m_{\pi_t} < m_t$, which is
disfavored by $R_b$ bound, the non $s$-channel contributions can
be quite large and dominant. (ii) Without the assumption
$m_{\pi_t}=m_{h_t}$, then for a relatively light top-Higgs, the
non $s$-channel contributions can be as large as $16\%$ for the
allowed range of $m_{\pi_t}$ (i.e. $>220$ GeV).

\begin{figure}
\begin{center}
\epsfig{file=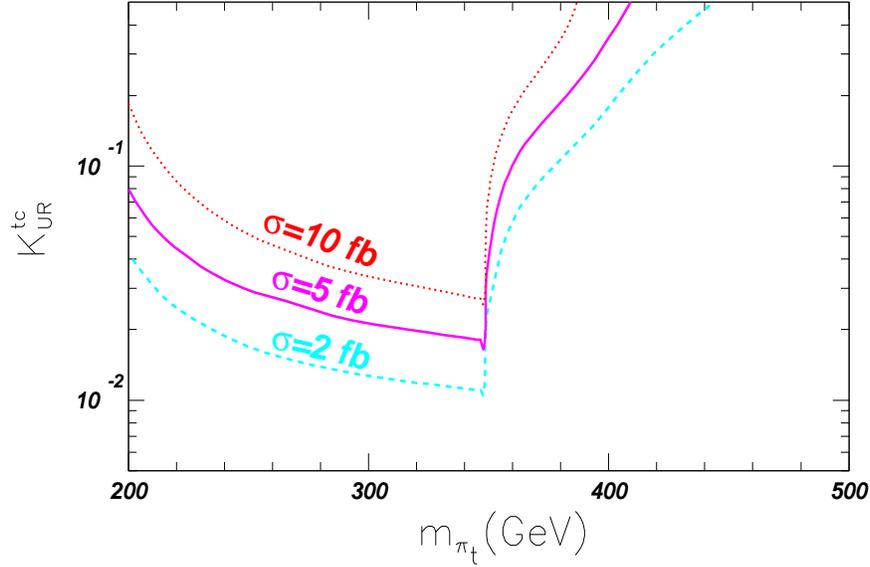,width=12cm,height=9.3cm} \caption {Contours of
the  cross section of $\gamma \gamma \to t \bar{c}$ in the plane
of $K_{UR}^{tc}$ versus $m_{\pi_t}$ for $\sqrt{s_{e^+ e^-}} =500$
GeV.} \label{fig:fig4}
\end{center}
\end{figure}

In Fig.\ref{fig:fig4} we plot the contours of the cross section in
the plane of $K_{UR}^{tc}$ versus $m_{\pi_t}$  under the
assumption $m_{h_t}=m_{\pi_t}$. One can learn that as
long as $m_{\pi_t}$ is lower than $400$ GeV, there
exists a large parameter space where the cross section can exceed
$2$ fb. Especially for $m_{\pi_t} \simeq 2 m_t $,
the parameter $K_{UR}^{tc}$ can be explored to as small as $10^{-2}$
given that a production rate $\sigma \simeq 2$ fb is accessible
at the ILC.

\begin{figure}
\begin{center}
\epsfig{file=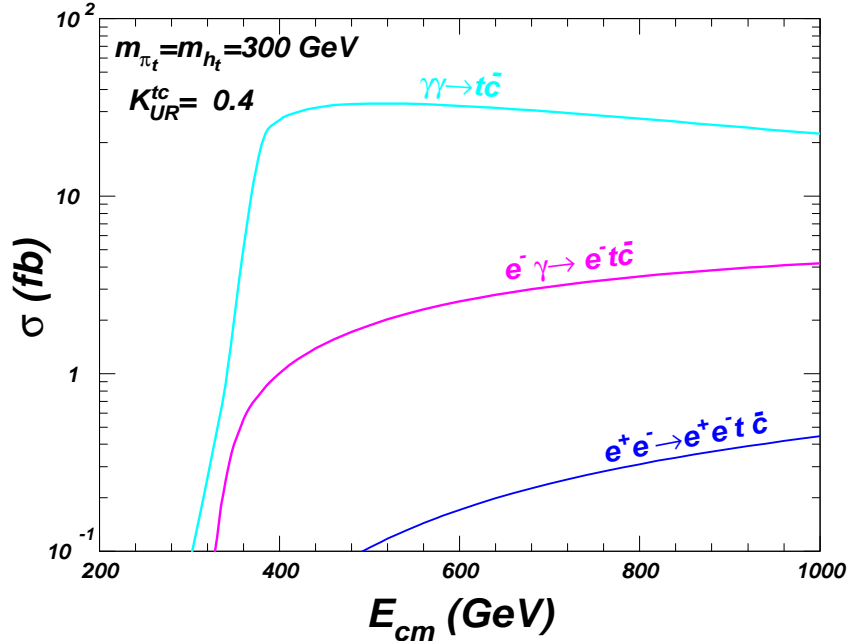,width=12cm,height=9cm} \caption {The cross
sections versus the $e^+e^-$ center-of-mass energy.}
\label{fig:fig5}
\end{center}
\end{figure}

In Fig.\ref{fig:fig5}, we show the behavior of the rates for
$\gamma \gamma \to t \bar{c} $, $ e^-\gamma \to e^- t \bar{c} $
and $ e^+ e^- \stackrel{\gamma^\ast \gamma^\ast}{\to} e^+ e^- t
\bar{c} $ versus the collider energy. As illustrated in the
figure, the $\sigma(\gamma \gamma \to t \bar{c})$ is insensitive
to the collider energy for $E_{cm} > 400$ GeV while the other two
channels rise significantly with the collider energy, which can be
explained by the energy dependence of $P_{\gamma/e}$.

Fig.\ref{fig:fig5} shows that $\sigma(\gamma \gamma \to t \bar{c})
\gg \sigma(e^- \gamma \to e^- t \bar{c}) \gg
\sigma(e^+ e^- \stackrel{\gamma^\ast \gamma^\ast}{\to} e^+ e^- t \bar{c})$.
Since the $s$-channel process $e^+ e^- \to \gamma^\ast,Z^\ast \to t \bar{c}$
is suppressed by the propagators of the intermediate photon or $Z$-boson
and was found  \cite{eetc-tc2} to occur at a similar rate as
$e^+ e^- \stackrel{\gamma^\ast \gamma^\ast}{\to} e^+ e^- t \bar{c}$
shown in Fig.\ref{fig:fig5},
we conclude that the production rate in $\gamma \gamma$ collision
is largest at a linear collider.


\section{Comparison of the predictions of different models}

In this section, we first briefly recapitulate the sources of FCNC transitions
in different models and then compare the typical magnitudes of various FCNC processes
at the ILC predicted by different models.
We arrive at the observation that the TC2 model predicts
much larger FCNC transitions than other new physics models.

It is well known that in the SM the FCNC transitions are absent at
tree-level and can occur only at loop level by the GIM mechanism.
The source of such FCNC transitions is the non-diagonality of the
Cabbibo-Kobayashi-Maskawa (CKM) matrix. In the extensions of the
SM, although the CKM  matrix can still induce additional
contributions to FCNC via loops composed by new particles, some
new sources for FCNC transitions usually come in.

As the simplest extension of the SM, the two-Higgs-doublet model
(2HDM) may naturally have FCNC mediated by the Higgs bosons at
tree-level, unless some ad hoc discrete symmetry is imposed. The
generic type of 2HDM is the so-called "type-III" model (2HDM-III)
\cite{2hdm,2hdm5}, in which the up-type and down-type quarks
couple to both Higgs doublets and thus the diagonalization of the
quark mass matrices does not automatically ensure the
diagonalization of the Yukawa couplings. In a popular realization
of 2HDM-III, one Higgs doublet is responsible for the electroweak
symmetry breaking as well as generating the fermion masses while
the other doublet has FCNC couplings whose strength are usually
parameterized as  \cite{2hdm}
\begin{eqnarray}
\xi_{ij}^{U,D}=\lambda_{ij}^{U,D} \frac{\sqrt{m_i m_j}}{v}
\label{coupling}
\end{eqnarray}
with $i,j=1,2,3 $ being the generation indices, $m_{i}$ the quark
masses and $\lambda_{ij}$ the dimensionless parameters. From
Eq.(\ref{coupling}), we can learn the FCNC coupling without top
quark is generally suppressed by the involved  quark masses. One
impressive feature of 2HDM-III is that the coupling strengths of
the FCNC are related with those of flavor-changing charged
currents. The FCNC couplings $\lambda_{tc}^U $ and
$\lambda_{ct}^U$ then contribute to some low energy observable and
the bounds from low energy data are $\lambda_{tc}^U $,
$\lambda_{ct}^U \sim O (1)$\cite{2hdm5}.

The MSSM \cite{MSSM} also contains
two Higgs doublets. However, the homology of the superpotential
requires that one doublet couple only to up-type quarks while the
other doublet couple only to down-type quarks and hence avoids
tree-level FCNC transitions in the Higgs sector. Note that when
supersymmetry is broken, the couplings $QUH_d$ and $Q D H_u$ are
generated at one-loop level to induce FCNC Yukawa couplings. While
such FCNC couplings are generally small  for up-type quarks, they
may be quite large for down-type quarks due to the enhancement by
large $\tan \beta $\cite{hall}. Another source of FCNC in the MSSM
is the flavor mixings of sfermions\cite{MSSM,susyflavor}, which
induce FCNC transitions in the fermion sector through loops
composed by sparticles, such as sfermion-gaugino loop and
sfermion-Higgsino loop\cite{tcv}.

As discussed in Sec. II, TC2 model, as one of the dynamic EWSB
models, is quite different from the 2HDM and the MSSM. In this
model, the scalar bosons are composite particles with properties
quite similar to those of Higgs bosons. One distinguished
character of TC2 model is that the third generation quarks are
special and have new top-color interaction\cite{tc2-Hill}. This
non-universal topcolor interaction can result in FCNC transitions
when expressing the interactions in terms of quark mass
eigenstates. In particular, the triangle texture of the up-type
quark mass matrix due to the generic top-color breaking pattern
can lead to large flavor mixing between $t_R$ and $c_R$
\cite{hjhe,top-Higgs}. Furthermore, the FCNC transitions also
exhibit themselves in the neutral scalar sector, which are $t
\bar{t} $ condensates. Therefore, the top flavor phenomenology is
much richer in this model than in other models.

So far as the top-charm associated productions are concerned, in
the SM they are severely suppressed since they proceed through
loops comprising of light down-type quarks (much lighter than top
quark mass scale) and involving the small mixings between the
third-generation quarks and the quarks of the first two
generations. Table I shows the rates for $e^+ e^- \to t \bar{c}$,
$e\gamma \to e t \bar{c}$ and $\gamma \gamma \to t \bar{c}$ at the
ILC.  As can be seen, the rates predicted by the SM are too small
to be accessible at the ILC.

In 2HDM-III, the process $\gamma \gamma \to t \bar{c}$ proceeds in
a similar way to that in TC2 model except that the top-pions and
top-Higgs in Fig.~\ref{fig:fig1} should be replaced by the
corresponding Higgs bosons. However, due to the smallness of
$\lambda_{tc}^U$, its cross section is generally two orders lower
than in the TC2 model. In Table I, we present the predictions of
2HDM-III for various production channels at the ILC. One
impressive character is that the rates of $e^+ e^-
\stackrel{Z^\ast Z^\ast}{\to} e^+ e^- t \bar{c}$ and $ e^+ e^-
\stackrel{W^\ast W^\ast}{\to} \nu \bar{\nu} t \bar{c} $ is
comparable to $\gamma \gamma \to t \bar{c}$ and much larger than
$e^+e^- \to t \bar{c}$. The reason is that in 2HDM-III the Higgs
bosons may couple at tree-level with $ZZ$ or $W^+ W^-$ and,
consequently, despite of the suppression of the probability to
find the gauge bosons in an electron, its cross section may still
be large\cite{2hdm3}.

In the MSSM, $\gamma\gamma \to t \bar{c}$ proceeds via the loops
comprising of squarks, gluinos, charginos and neutralinos. The
corresponding results in Table I are taken from \cite{eetc-mssm}
where only SUSY-QCD contributions are considered \footnote{To our
knowledge, the SUSY electroweak contributions to top-charm
associated production have not yet been calculated. However, from
the fact that the SUSY-QCD contributions to the rare decay $t \to
c \gamma$ is generally larger than SUSY electroweak contributions
\cite{tcv}, we may infer that the SUSY-QCD contributions represent
the typical size of SUSY contributions.}. As shown in Table I, the
cross sections in the MSSM, although can be much large than the SM
predictions, are generally much smaller than the TC2 predictions.
The reason is twofold. One is that in the MSSM there is no
$s$-channel contribution like that shown in Fig.1 (r) at one-loop
level. The other reason is that there are cancellations between
different diagrams due to the unitarity of the matrix
diagonalizing the up-type squark mass matrix, which is often
called the 'Super-GIM' mechanism.

\null \noindent {\small Table I: Theoretical predictions for
top-quark FCNC processes. The predictions of new physics models
are optimum values. The $e^+e^-$ collider energy is 500 GeV for
production processes. The cross sections are in the units of fb.
The processes not referenced are estimated by us. }
\vspace*{0.1cm}
\begin{center}
\begin{tabular}{lllll}
\hline
& ~~~SM  & ~~~2HDM-III & ~~~MSSM & ~~~TC2 \\
 \hline $ \sigma (\gamma \gamma \to t \bar{c})\ \ \  $ &\ \ \  ${\cal{O}}(10^{-8}) $ \cite{eetc-mssm}\ \ \   &
\ \ \  ${\cal{O}}(10^{-1}) $
\cite{ma2}\ \ \    & \ \ \  ${\cal{O}}(10^{-1}) $ \cite{eetc-mssm}\  \ \ &\  \ \ ${\cal{O}}(10) $ \ \ \ \\
\hline
 $ \sigma (e \gamma \to e t \bar{c})  $\ \ \ & \ \ \  ${\cal{O}}(10^{-9}) $   \cite{eetc-mssm}\ \ \ & \ \ \  $
 {\cal{O}}(10^{-2}) $ \ \ \    &\ \ \ $ {\cal{O}}(10^{-2}) $ \cite{eetc-mssm}\  \ \ & \  \ \ ${\cal{O}}(1) $  \ \ \ \\ \hline
 $ \sigma (e^+ e^- \to t \bar{c}) $\ \ \  &\ \ \  ${\cal{O}}(10^{-10}) $  \cite{sm2,eetc-mssm} \ \ \
 &  \ \ \  ${\cal{O}}(10^{-3}) $ \cite{2hdm2}\ \ \   & \ \ \ $ {\cal{O}}(10^{-2}) $
\cite{eetc-mssm}\ \ \ & \  \ \ ${\cal{O}}(10^{-1}) $
\cite{eetc-tc2}\ \ \ \\ \hline
 $ \sigma (e^+ e^- \stackrel{\gamma^\ast \gamma^\ast}{\to} e^+ e^-  t \bar{c}) $\ \ \  &\ \ \  $< 10^{-10} $  \cite{eetc-mssm} \ \ \
  & \ \ \  ${\cal{O}}(10^{-3}) $  \ \ \  & \ \ \  ${\cal{O}}(10^{-3}) $  \cite{eetc-mssm} \ \ \  &
   \ \ \  ${\cal{O}}(10^{-1}) $  \ \ \  \\ \hline
 $  \sigma (e^+ e^- \stackrel{Z^\ast Z^\ast}{\to} e^+ e^-  t \bar{c})$\ \ \ & \ \ \  $< 10^{-10} $   \ \ \ &
  \ \ \  ${\cal{O}}(10^{-1}) $  \cite{2hdm3}  \ \ \   & \ \ \ $ < 10^{-3} $   &  \  \ \ ${\cal{O}}(1) $   \cite{ee-ht}\ \ \ \\ \hline
 $  \sigma (e^+ e^- \stackrel{\gamma^\ast Z^\ast}{\to} e^+ e^-  t \bar{c}) $ \ \ \ & \ \ \  $ < 10^{-10} $  \ \ \ &
\ \ \  $ < 10^{-3} $  \ \ \   & \ \ \ $ < 10^{-3} $   & \  \ \ $<
10^{-1} $  \ \ \ \\ \hline
 $  \sigma (e^+ e^- \stackrel{W^\ast W^\ast}{\to} \nu \bar{\nu}  t \bar{c}) $ \ \ \ & \ \ \  $ < 10^{-10} $  \ \ \ &
 \ \ \  ${\cal{O}}(10^{-1}) $  \cite{2hdm3}  \ \ \ &  \ \ \ $ < 10^{-3} $  & \  \ \ ${\cal{O}}(1) $  \cite{ee-ht} \ \ \ \\ \hline
 $  Br( t \to c g) $\ \ \  & \ \ \ ${\cal{O}}(10^{-11}) $ \cite{sm3} \ \ \ &\ \ \ ${\cal{O}}(10^{-5}) $ \cite{2hdm4}\ \ \ &
 \ \ \ ${\cal{O}}(10^{-5}) $ \cite{tcv,eetc-mssm}\ \ \   & \ \ \ ${\cal{O}}(10^{-4}) $ \cite{tctcv}\ \ \  \\ \hline
 $  Br( t \to c Z)  $ \ \ \ & \ \ \ ${\cal{O}}(10^{-13}) $ \cite{sm3}\ \ \ & \ \ \ ${\cal{O}}(10^{-6}) $ \cite{2hdm4} \ \ \ &
  \ \ \ ${\cal{O}}(10^{-7}) $ \cite{tcv,eetc-mssm}\ \ \   &  \ \ \ ${\cal{O}}(10^{-4}) $ \cite{tctcv}\ \ \ \\ \hline
 $  Br( t \to c \gamma) $ \ \ \ & \ \ \ ${\cal{O}}(10^{-13}) $ \cite{sm3} \ \ \ & \ \ \ ${\cal{O}}(10^{-7}) $ \cite{2hdm4}
  \ \ \ &  \ \ \ ${\cal{O}}(10^{-7}) $ \cite{tcv,eetc-mssm}\ \ \   & \ \ \ ${\cal{O}}(10^{-6}) $ \cite{tctcv}\ \ \ \\ \hline
$  Br( t \to c h) $ \ \ \ & \ \ \ $< 10^{-13} $ \cite{sm3} \ \ \ &
\ \ \ ${\cal{O}}(10^{-3}) $ \ \ \ &  \ \ \ ${\cal{O}}(10^{-4}) $
\cite{MSSM3}\ \ \  & \ \ \ ${\cal{O}}(10^{-1}) $  \ \ \ \\ \hline
\end{tabular}
\end{center}

For completeness we also present the branching ratios of various
top rare decays in Table I. It may be surprising that the
branching ratio of $t \to c h_t$ can reach $10^{-1}$ in TC2 model
if the top-Higgs is light. In fact, this does not contradict with
current experiments due to the small statistics of current top
quark measurements \cite{review,top}. Future bound on $t \to c h_t
$ will not influence the optimum magnitude of $ \gamma \gamma \to
t \bar{c} $ since the favored region for the latter process is $
m_{h_t}= 200 \sim 300 $ GeV.

We conclude from Table I that TC2 model generally predicts much
larger top quark FCNC transitions than any other models and the
$\gamma \gamma$ collision is the best channel in enhancing the
magnitude for the top-charm associated productions at the ILC.

\section{Observability of top-charm production at ILC}

Given the predictions listed in Table I, we now discuss their
observability at the ILC. First, for the top rare decays other
than $t \to c h$, the hope to observe them at the ILC is dim since
only about $10^4$ top quark events can be produced for an
integrated luminosity of $100$ fb$^{-1}$ at the ILC
\cite{review,bran}. For the top-charm associated production
processes, the observability is analyzed in the following.

The cleanest signal for top-charm associated productions
at the ILC is $\ell bj+\p_slash_T$ with $j$ being a light quark
jet and $\ell =e $ or $\mu $. Generally speaking, the SM irreducible
backgrounds for these processes are small due to the odd $b$-parity
of the signal\cite{background}.
So far as $e^+ e^- \to t \bar{c}$ is concerned, the irreducible SM
background arises from $ e^+ e^- \to W^+ W^- \to \bar{c} b l \nu$
and is negligible due to the small size of $V_{cb}$.
The leading SM background then comes from
\begin{eqnarray}
e^+ e^- \to q \bar{q}^{\prime} l \nu
\end{eqnarray}
where the light quark jet $q$ or $q^\prime$ may be mis-identified as
a $b$-jet. Such backgrounds, mainly from $W$ pair productions
as well as the $W$ bremsstrahlung processes $e^+e^- \to W +2~jets$,
can reach $2252$ fb in total \cite{x3}.
Fortunately, these backgrounds can be efficiently suppressed by
reconstructing the top-quark mass from the c.m. energy and
the charm jet energy \cite{x3}, {\it i.e.}
\begin{eqnarray}
m_t^{rec} =(s-2 \sqrt{s} E_c )^{1/2} \label{reconstruct}
\end{eqnarray}
with $E_c =\frac{\sqrt{s}}{2} (1- m_t^2/s)$. According to the
analysis in Ref.~\cite{x3}, the $t\bar c$ production with a cross
section larger than $1$ fb is observable at $95\%$ C.L.  for an
integrated luminosity of $100$ fb$^{-1}$. From Table I one sees
that no new physics models can enhance the rate of $e^+ e^-\to t
\bar{c}$ to the level of  $1$ fb.

Next we turn to the process $e^+ e^- \stackrel{W^\ast W^\ast}{\to}
\nu \bar{\nu} t \bar{c} $. The signal is not as distinctive as
$e^+ e^- \to t \bar{c}$ due to the missing neutrinos. Its SM
reducible background is generated by processes such as $e^+ e^-
\to W^+ W^- \nu_e \bar{\nu}_e $ and $ e^+ e^- \to t \bar{t} \nu
\bar{\nu}_e$. These backgrounds can be suppressed by applying some
useful cuts \cite{x3}. So far the detailed Monte Carlo simulation
for the observability of this process is still lacking. We expect
conservatively that the production at the level of several fb may
be accessible at the ILC. From Table I we see that TC2 model can
enhance this production to this level.

Finally, we consider the $\gamma \gamma$ collision. The largest SM
background for $\gamma \gamma \to t \bar{c}$ is from $\gamma
\gamma \to W^+ W^-$, which can reach ${\cal{O}}(10)$ pb. Assuming
a fixed c.m. energy of 500 GeV for $\gamma \gamma$ collision, a
detailed Monte Carlo simulation \cite{effect2} showed that the
background can be neglected at the expense of reducing the signal
cross section to $14 \%$. Noting the fact that the cuts used in
\cite{effect2} are not sensitive to the energy of $\gamma\gamma$
collision, one may infer that this conclusion is approximately
valid for a realistic $\gamma \gamma$ collision whose c.m. energy
is not fixed. In fact, this point was also emphasized at the end
of Sec. III and Sec. IV in Ref.~\cite{effect2}. In practice, if we
assume conservatively that the signal is reduced to $10 \%$ to
eliminate backgrounds, we may expect that the production $\gamma
\gamma \to t \bar{c}$ as large as $5$ fb may be accessible at the
ILC at $3 \sigma $ level. Compared with the predictions in Table
I, one sees that TC2 model can enhance the production $\gamma
\gamma \to t \bar{c}$ to the observable level at the ILC in a
large part of the parameter space.

\section{ Summary and Conclusion}

We calculated the top-charm associated productions via $e^+ e^-$,
$e^- \gamma$ and $\gamma \gamma$ collisions at linear colliders in
the topcolor-assisted technicolor model. Then we compared the
results with the existing predictions of the SM, the general
two-Higgs-doublet model and the Minimal Supersymmetric Model. We
observed that the topcolor-assisted technicolor model predicts
much larger production rates than other models and the
largest-rate channel is $\gamma \gamma \to t \bar{c}$, which can
exceed 10 fb for a large part of the parameter space. From the
analysis of the observability of such productions at the
future linear colliders, we conclude that the predictions
of the topcolor-assisted technicolor model can reach the
observable level for a large part of the parameter space while the
optimum predictions of other models may lie below the accessible
level.

\section*{Acknowledgment}
This work is supported in part by Young Outstanding Foundation of
Academia Sinica, by Chinese Natural Science Foundation under No.
10175017, 10375017 and by the Invitation Program of JSPS under No.
L03517.

\end{document}